\documentclass{epl}

\title{Unusual interplay between copper-spin and vortex\\ dynamics in 
slightly overdoped \chem{La_{1.83}Sr_{0.17}CuO_4} }
\author{R. Gilardi\inst{1}, A. Hiess\inst{2}, N. Momono\inst{3}, M. Oda\inst{3}, M. Ido\inst{3} \and J. Mesot\inst{1}}
\institute{
  \inst{1} Laboratory for Neutron Scattering, ETH Zurich and PSI 
             Villigen - CH-5232 Villigen, Switzerland\\
  \inst{2} Institute Laue-Langevin - BP 156, F-38042 Grenoble Cedex\\
\inst{3} Departement of Physics, Hokkaido University - Sapporo 
         060-0810, Japan
}

\begin{document}

\maketitle

\begin{abstract}
Our inelastic neutron scattering experiments of the spin excitations 
in the slightly overdoped  La$_{1.83}$Sr$_{0.17}$CuO$_{4}$ compound 
show that,
under the application of a magnetic field of 5 Tesla, 
the low-temperature susceptibility undergoes a weight redistribution 
centered at the spin-gap energy. Furthermore, by comparing the 
temperature dependence of the neutron data 
with ac-susceptibility and magnetization measurements, we conclude that 
the filling in 
of the spin gap tracks the irreversibility/melting temperature 
rather than $T_{c2}$, which indicates an unusual interplay between the 
magnetic vortices and the spin excitations even in the slightly overdoped regime of high-temperature superconductors.
\end{abstract}

\section{Introduction}
Experiments on the high-temperature superconductors (HTSC) have 
produced a 
tremendously rich variety of physical behaviours, both from the mesoscopic (physics of vortices \cite{BLATTER,Giamarchi,Landau}) as well as from the 
microscopic (fundamental electronic \cite{Norman} and magnetic excitations \cite{Kastner}) point of views.

From the \textit{mesoscopic} point of view, the cuprates are type-II 
superconductors with strong anisotropy giving rise to a complicated 
magnetic phase diagram \cite{BLATTER}: below the lower critical field 
$H_{c1}$, 
the magnetic flux is 
entirely excluded from the material (Meissner phase); above $H_{c1}$ 
quantized magnetic vortices can penetrate the superconductor 
(mixed phase) to form an ordered vortex lattice that eventually melts 
at higher fields
and temperatures, giving rise to a vortex fluid phase; finally above 
the upper 
critical field $H_{c2}$ the 
normal state is recovered.
In La$_{2-x}$Sr$_{x}$CuO$_{4}$ (LSCO) the first order melting transition is almost concomitant to the 
irreversibility line $H_{irr}$, where reversible magnetization and
resistivity appear \cite{SASAGAWA, Gilardi2}. 
Surprisingly, very little information exists about the microscopic 
observation of 
vortex lattice in LSCO,
a compound belonging to the family of the first HTSC that has been
discovered. Using small angle neutron 
scattering (SANS), we have recently reported the first direct 
evidence of 
a well ordered vortex lattice in slightly overdoped LSCO (x=0.17), which furthermore undergoes a field-induced hexagonal to square transition at around 0.4 Tesla \cite{GILARDI,Gilardi2}.

From the \textit{microscopic} point of view, the 
LSCO compounds are 
characterized by the presence of spin excitations 
located in the vicinity of the antiferromagnetic (AF)
wavevector of the undoped parent compound at the 
$(\frac{1}{2}\pm\delta,\frac{1}{2})$ and 
$(\frac{1}{2},\frac{1}{2}\pm\delta)$ incommensurate 
positions \cite{Cheong}.
While such excitations could originate from Fermi surface nesting 
properties due to 
coherence effects in the superconducting state \cite{BULUT,Norman2}, it has 
also been proposed
that they could be the signature of either dynamical 
stripes \cite{TRANQUADA}
or SO(5) supersymmetry \cite{ZHANG}.
Since these various models differ in the way the AF state of the undoped 
compound is related to the superconducting state, the application of a magnetic field provides a 
clean way to differentiate between them.
While conventional vortices emerge from Fermi-liquid like models,
the SO(5) model predicts unusual vortices with bound AF-states 
\cite{Hu}. To our knowledge, no predictions exist for the 
stripes-model in a magnetic field.

Recently, Lake \textit{et al.} \cite{LAKE01} have reported the 
magnetic field dependence of the incommensurate spin excitations in 
 \textit{optimally} doped LSCO. On one side they observe, below $T$=10 
K, the appearence of 
sub-gap excitations when a field of 7.5 Tesla is applied parallel to 
the c-axis. On the other side it is 
shown that the filling in of the gap with increasing temperature 
seems to track the 
irreversibility/melting line rather than $T_{c2}$. Both observations 
are 
highly unusual and it is crucial to understand whether or not they share a 
common origin. In order to answer this question, we have performed 
field dependent measurements of the incommensurate excitations 
on a slightly  \textit{overdoped} LSCO crystal. 

\section{Experimental results}
Our experiments were performed on the same high-quality single crystal as the one used 
for the vortex lattice measurements \cite{GILARDI}. Details of the 
sample growth can be found 
elsewhere \cite{ODA}. The neutron data were taken on the IN22 
spectrometer at the high-flux reactor of the Institute Laue Langevin 
in Grenoble, France. We used a graphite vertically curved 
monochromator 
and graphite horizontally curved analyser 
with a fixed final energy of 13.8 meV. In order to avoid 
contamination by higher-order reflections a PG-filter in $\mathbf{k_{f}}$ was installed.
The sample, a LSCO single crystal (x=0.17, $T_{c}$=37 
K, $\Delta T_c$=1.5 K, $m$=2727~mg), was mounted in a cryostat with the c-axis oriented perpendicular 
to the 
scattering plane. A magnetic field up to 5~Tesla could be applied parallel 
to 
the c-axis. The \textbf{Q}-scans were performed by rotating the 
sample around 
its c-axis. Within such a scan, one can 
measure two different incommensurate peaks without changing the 
analyser and detector positions.

The magnetic cross section for inelastic neutron scattering is given 
by
\begin{equation}
    \frac{d^{2}\sigma}{d\Omega d\omega} \sim 
   \frac{k_{f}}{k_{i}}f^{2}(\mathbf{Q})
\sum_{\alpha,\beta}^{}(\delta_{\alpha\beta}-\frac{Q_{\alpha}Q_{\beta}}{Q^{2}})
S^{\alpha\beta}(\mathbf{Q},\omega)
   \label{eq:magnetisch}
\end{equation}
where $f(\mathbf{Q})$ is the magnetic form factor, 
$S^{\alpha\beta}(\mathbf{Q},\omega)$ is the dynamic structure 
factor, $\hbar\omega=E_{i}-E_{f}$ and 
$\mathbf{Q}=\mathbf{k_{i}}-\mathbf{k_{f}}$ are the neutron energy and 
momentum transfers.
The imaginary part of the generalized susceptibility is related to 
the 
measured dynamic structure factor via the fluctuation-dissipation 
theorem,
\begin{equation}
    \chi''(\mathbf{Q},\omega) = 
    S(\mathbf{Q},\omega)(1-e^{\frac{-\hbar\omega}{k_{B}T}})
   \label{eq:fluct-diss}
\end{equation}

\begin{figure}
\onefigure[scale=0.5]{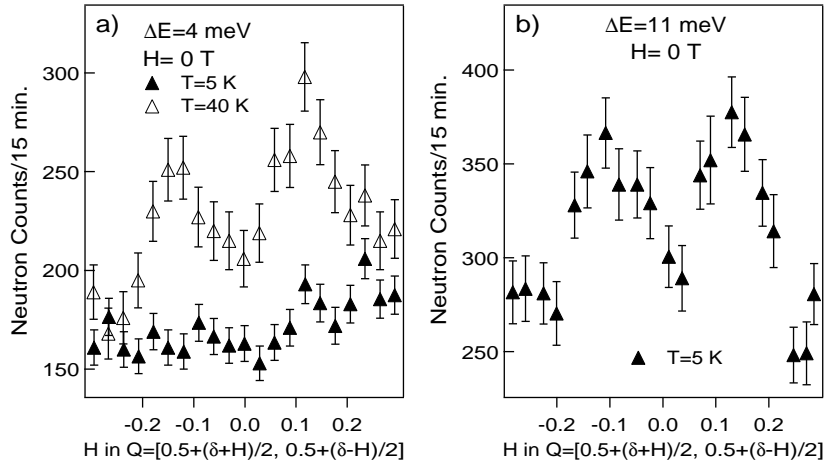}
\caption{Constant energy-scans through two incommensurate peaks at 
zero field: a) for an energy transfer 
$\hbar\omega$=4~meV, $T$=5~K and 40~K, and b) for $\hbar\omega$=11~meV, 
$T$=5~K. }
\label{fig1}
\end{figure}

We now start with the description of the data obtained at zero-field.
Fig.$\ref{fig1}$a shows \textbf{Q}-scans through the incommensurate 
peaks at  an energy transfer ${\hbar\omega}$=4~meV for temperatures above and 
below $T_{c}$. 
Fig.$\ref{fig1}$b shows a similar scan at ${\hbar\omega}$=11~meV and 
T=5~K.
Above $T_{c}$ we observe clear peaks at the expected incommensurate positions 
$\mathbf{Q_{\delta}}=(\frac{1}{2},\frac{1}{2}+\delta)$ and 
$(\frac{1}{2}+\delta,\frac{1}{2})$
with $\delta$=0.13(1) \cite{YAMADA}. Below $T_{c}$ we 
observe that the magnetic intensity at low energy disappears, which 
is a clear sign for the opening of a spin gap in the superconducting 
state.

\begin{figure}
\onefigure[scale=0.5]{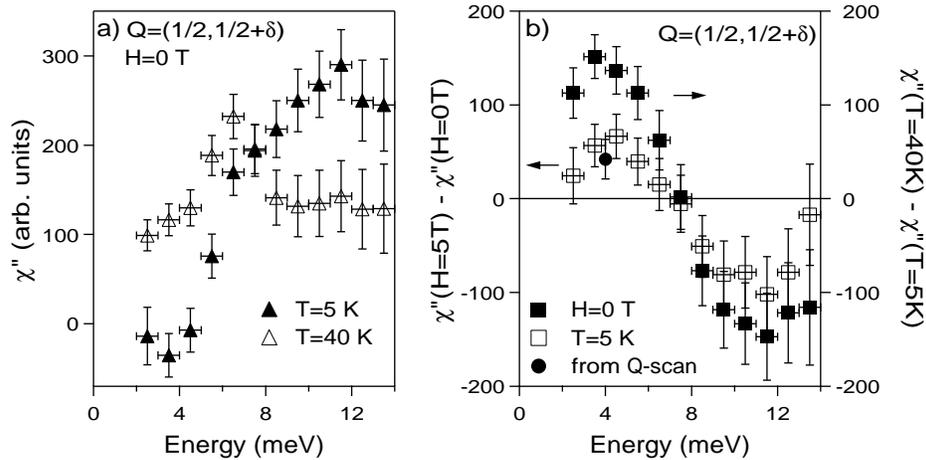}
\caption{a) Energy dependence of  $\chi''(\mathbf{Q_{\delta}},\omega)$ in zero-field at 
$T$=5~K and 40~K. A background 
measured away from $\mathbf{Q_{\delta}}$ has 
been subtracted from the raw neutron data. While at high 
temperature the intensity is roughly constant, at low temperature a 
gap is present. 
b) Difference of the high- and 
low-temperature susceptibility $\chi''(\mathbf{Q_{\delta}},\omega)$ (shown in Fig.$\ref{fig2}$a) measured at $H$=0 T, and of the $H$=5 T and zero-field susceptibility $\chi''(\mathbf{Q_{\delta}},\omega)$ (shown in Fig.$\ref{fig3}$b and Fig.$\ref{fig2}$a) measured at T=5 K. Similarly to the temperature effect, the application of a magnetic 
field induces a transfer of weight from high- to low-energies. The field-induced increase of $\chi''$ at 4 meV, obtained independently from the \textbf{Q}-scan (see text), has also been included (filled circle).}
\label{fig2}
\end{figure}

\begin{figure}
\onefigure[scale=0.5]{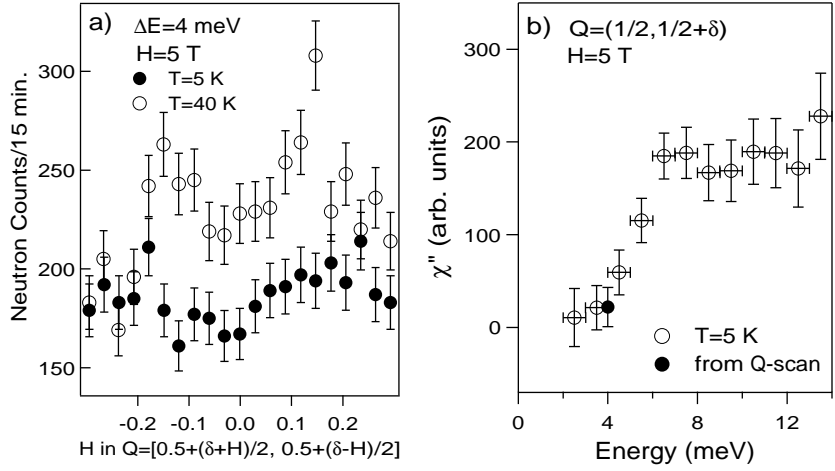}
\caption{a) Constant energy-scans through two incommensurate peaks in a field of $H$=5 T for $\hbar\omega$=4 meV, $T$=5~K and 40~K. b) Energy dependence of  $\chi''(\mathbf{Q_{\delta}},\omega)$ at 
$T$=5~K and $H$=5 T. The filled circle indicates the value of $\chi''$ at 4 meV obtained independently from the \textbf{Q}-scan measured at H=5 T, see text.}
\label{fig3}
\end{figure}

The presence of the gap is further confirmed by the energy scans 
performed at $\mathbf{Q_{\delta}}$, see Fig.$\ref{fig2}$a. 
For $T=5 K<T_{c}$,  $\chi''(\mathbf{Q_{\delta}},\omega)$ 
drops sharply below about 6.5 meV, which we identify with the spin gap 
energy $\Delta_{SG}$, whereas for $T=40K>T_{c}$, $\chi''(\mathbf{Q_{\delta}},\omega)$ 
doesn't vary much as a function of energy.

We now turn to the magnetic field dependence of the spin excitations.
Both \textbf{Q}- and energy scans are not strongly affected by the 
application of an external magnetic field 
of 5 Tesla applied perpendicular to the CuO$_{2}$ planes, as shown
in Fig.$\ref{fig3}$. In particular, in our slightly overdoped sample, there is no clear indication of field-induced
sub-gap excitations at low temperatures.
However one can notice that in the \textbf{Q}-scan performed at $H$=5 T ($T$=5 K, Fig.$\ref{fig3}$a) there is slightly more intensity than in the 
\textbf{Q}-scan performed at $H$=0 T ($T$=5 K, Fig.$\ref{fig1}$a). By taking the zero-field data at $T$=5 K as background and fitting two Gaussians with fixed positions (H=$\pm\delta$) and widths, we estimate that the value of $\chi''(H$=5 T, $T$=5 K$)$ is about 30$\%\pm15\%$ of $\chi''(H$=0 T, $T$=40 K$)$. This  value is in agreement with $\chi''(\mathbf{Q_{\delta}},\omega)$ obtained from energy scan, and is included in Fig.$\ref{fig2}$b and Fig.$\ref{fig3}$b (filled circle).
In a first step we have analyzed the energy dependence of the susceptibility using 
the phenomenological function introduced by Lee \textit{et al.} 
\cite{LEE2}. At low temperature and zero-field, we obtain a spin gap value 
of $\Delta_{SG}=6.5\pm 
0.4$ meV consistent with earlier results \cite{LEE2,LAKE99}. At 5 Tesla we observe a decrease of $\Delta_{SG}$ of the order of 25$\%$ ($\pm 20\%$). However, due to the large error on the determination of the spin gap value, no definitive conclusion can be reached.

It is very instructive to look on one side at the 
difference of the high- and low-temperature susceptibility $\chi''(\mathbf{Q_{\delta}},\omega)$ in zero-field, and on the other side at the difference 
between the 5 Tesla and zero-field susceptibility $\chi''(\mathbf{Q_{\delta}},\omega)$ at $T$=5 K (see Fig.$\ref{fig2}$b).  In both cases we observe a weight transfer from the 
high-energy to the low-energy region, 
most likely indicating a conservation of spin as reported 
earlier \cite{MASON} and also expected in more recent theoretical 
studies \cite{MORR}. 
The low-energy field-induced weight transfer at 5 Tesla is about 35$\%$ ($\pm15\%$) of the temperature-induced weight transfer 
measured between $T$=5 K and $T$=40K through $T_c$. This result is very 
surprising if one considers that the applied field represents only 
10$\%$ of $H_{c2}$(5K)$\sim$50 Tesla (as determined 
from resistivity data \cite{ANDO}).

\begin{figure}
\twofigures[height=8cm]{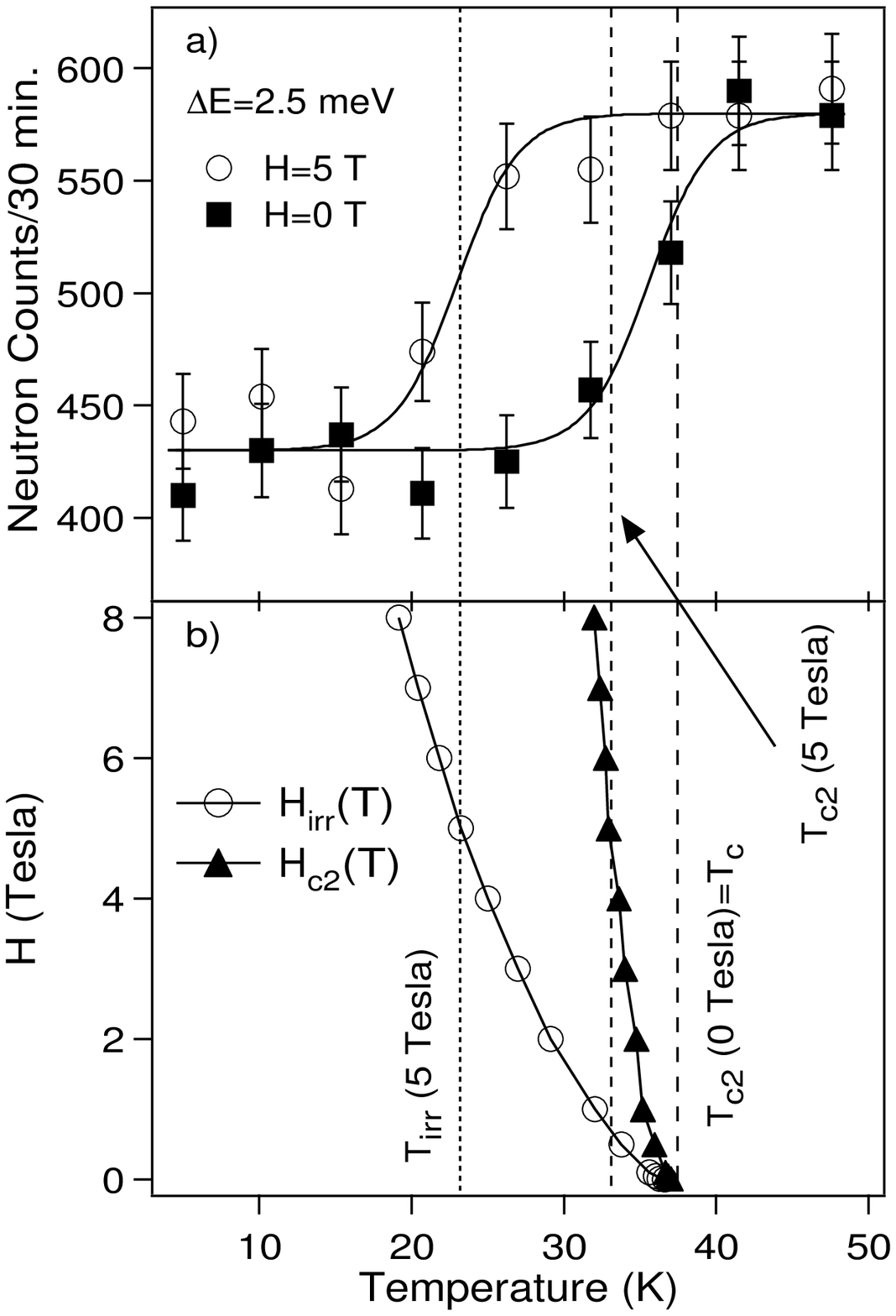}{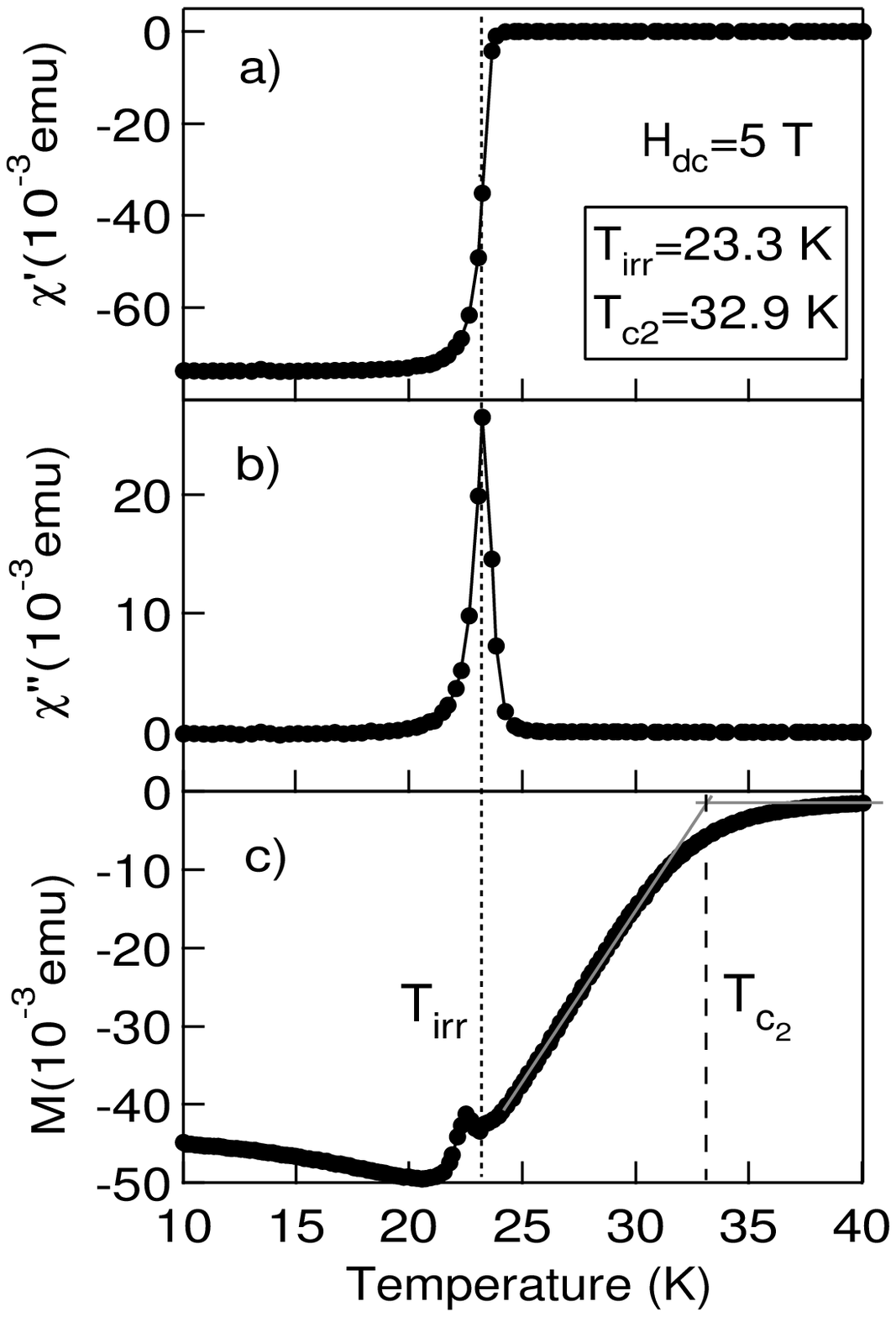}
\caption{a) Temperature dependence of the neutron intensity for an 
energy transfer $\hbar\omega$=2.5 meV taken in zero-field
and in a field of $H$=5 T (measured heating up after field-cooling). b) Irreversibility and 
superconducting transition lines 
 $H_{irr}$ and $H_{c2}$, respectively.
 Solid lines are guides to the eye, whereas dashed lines indicate the 
 values of $T_{c2}$ and $T_{irr}$ at $H$=0 T and $H$=5 T. The opening of the 
spin gap tracks $T_{irr}$ rather than 
$T_{c2}$ (see text).}
\label{Fig4}
\caption{a) Real and b) imaginary part of the ac-susceptibility 
$\chi=\chi'+i\chi''$ for $H=H_{ac}\cdot cos(\omega_{ac} t)+H_{dc}$ applied parallel 
to the c-axis with $H_{dc}$=5 T, $H_{ac}$=10 Oe and $\omega_{ac}$=10 Hz. 
$T_{irr}$ is given by the peak in $\chi''$  (or maximal slope of 
$\chi'$). c) Field-cooled magnetization in a dc-field of 5 T applied parallel 
to the c-axis . $T_{c2}$ is 
determined as explained in the text (extrapolation method). Notice 
that even in $M(T)$ there is a sign of the irreversibility line (dip on the 
right side of the small peak).}
\label{Fig5}
\end{figure}

Finally, we present the
temperature dependence of the spin fluctuations as a function of 
magnetic 
field. Fig.$\ref{Fig4}$a shows the neutron counts (without background 
subtraction) as a function of temperature at 
$\mathbf{Q_{\delta}}$
and at an energy transfer of 2.5 meV. The background was found to be 
dependent on the energy transfer, but independent on temperature ($T<40$ K) and 
magnetic field ($H<5$ T). The sharp decrease of the intensity by 
decreasing temperature indicates the opening of the spin gap.  At 
zero field the intensity drops around $T_{c}$, whereas at 5 Tesla the 
point at which the gap opens is shifted by about 15~K toward 
lower temperature. Similar results have been obtained for an energy 
transfer of 4 meV.
This indicates an unusually strong effect of 
the magnetic field on the temperature dependence of the spin 
fluctuations.

In order to understand this shift we performed 
ac-susceptibility ($\chi=\chi'+i \chi"$) and magnetization ($M$) measurements on a portion of the 
crystal used for the neutron experiments
\footnote{These measurements have been performed with a PPMS at the Paul 
Scherrer Institute, Switzerland. The whole data will be published 
elsewhere.}
The experimental irreversibility lines are obtained from the loss 
peak of the imaginary part $\chi''$ of the ac-susceptibility 
\cite{GOMORY}, which is 
directly related to the maximum slope of its real part $\chi'$. In 
Fig.$\ref{Fig5}$a+b we show the 
ac-susceptibility data for a dc-field of 5 Tesla applied parallel to the 
c-axis. The small 
width of the loss peak indicates the high quality of our 
single crystal.
The $H_{c2}(T)$ line is more difficult to measure, since it is 
only a crossover rather than a true phase transition.
Magnetization measurements have been used extensively to determine the 
upper critical field. The simplest approach is to use the linear 
extrapolation method based on the linear Abrikosov formula for 
magnetizations at high fields near $H_{c2}$ \cite{Abrikosov}. The 
transition 
temperature $T_{c2}$ is derived from the intersection of the linear 
fit with 
the normal-state horizontal line, as shown in Fig.$\ref{Fig5}$c.
It was argued
that this procedure is not totally correct for HTSC, where the 
Abrikosov linear region is limited to a small temperature range, because of the rounding of the $M(T)$ curves close to $T_{c2}$ (due to either diamagnetic fluctuations or sample inhomogeneities)\cite{HAO,OTT}. However, in our case, the $H_{c2}(T)$ line determined by extrapolation is in quite good agreement with the line obtained by scaling procedures \cite{OTT}.
By applying dc magnetic fields up to 8 T parallel to 
the c-axis we determined $T_{irr}$ and $T_{c2}$ and obtained the 
magnetic phase diagram shown in Fig.$\ref{Fig4}$b. 
Notice that between zero and 5 
Tesla, $T_{c2}$ changes by only 4 K. Therefore, one would expect 
that the application of 
such a magnetic field would only marginally affect the temperature at which the spin gap opens. As can be seen in Fig.$\ref{Fig4}$a, this is 
clearly not the case since in a field 
of 5 Tesla the point at which the gap starts to fill in is shifted 
toward a much lower temperature than the measured $T_{c2}$, and 
agrees well with the measured $T_{irr}$. Interestingly, in our SANS experiments \cite{Gilardi2}, we observed that the vortex lattice intensity vanishes exactly at the irreversibility line, whereas no indication of vortices can be observed above $T_{irr}$.

The suppression of the spin gap in the vortex fluid phase above $T_{irr}$ can be
understood qualitatively by assuming that in this region of the phase
diagram, the time scale characterizing  the dynamics of the vortices becomes comparable
to that of the Cu-spin excitations \cite{LAKE01}. Such a coupling between the electronic
and vortex-fluid degrees of freedom has been inferred earlier, on one hand
from Hall measurements where a sharp change of the conductivity is observed
at the melting transition \cite{DAnna}, and on the other hand from the
observation of an unusually large increase of the specific heat when passing
from the solid- to the fluid-vortex state \cite{Schilling}. All these experiments,
including our own, point toward a more complex nature of the vortex fluid
state than anticipated so far. It remains a theoretical challenge to
understand, at a quantitative level, the connection existing between these
anomalous observations.

\section{Conclusions}
In conclusion, although our measurements of the spin excitations in a 
slightly overdoped LSCO do not reveal clear evidence for
magnetic field induced sub-gap excitations, we do observe an 
unusually large spectral weight redistribution centered at the spin gap 
energy when a field of 5 Tesla is applied.
It is interesting to put our results in perspective with those obtained by Lake \textit{et al.} \cite{LAKE01,LAKE02}:
in the underdoped regime the magnetic 
field induces static $(\hbar\omega=0)$ moments \cite{LAKE02}, while at 
optimal doping excitations at finite energy but within the spin gap 
$(0<\hbar\omega<\Delta_{SG}$) are created \cite{LAKE01}. Our own results 
show that upon further doping, weight transfer 
centered at the the spin gap energy ($\hbar\omega \sim \Delta_{SG}$)  occurs.
This suggests the existence, beside the superconducting and spin gap energies, of 
an additional doping dependent magnetic energy scale.

Finally, by comparing the temperature dependence of the neutron data 
with macroscopic measurements, we conclude that the filling in 
of the spin gap tracks the irreversibility/melting temperature, 
rather than $T_{c2}$.
This indicates that even in the overdoped regime of HTSC there exists an unusual coupling between the  copper-spin and vortex degrees of freedom.

\acknowledgments
We thank L.-P. Regnault for his assistance during the experiment.
This work was supported by the Swiss National Science Foundation
and the Ministry of Education and Science of Japan.


\begin{thebibliography}{0}

\bibitem{BLATTER}
For a review, see \Name{G. Blatter \textit{et al}.}
\REVIEW{Rev. Mod. Phys.}{66}{1994}{1125}.

\bibitem{Giamarchi}
\Name{T. Giamarchi and S. Bhattacharya}
  \Book{in High Magnetic Fields: Applications in Condensed Matter Physics and Spectroscopy}
  \Editor{C. Berthier \textit{et al.} }
  \Publ{Springer-Verlag}
  \Year{2002}
  \Page{314};\\ cond-mat/0111052.

\bibitem{Landau}
\Name{I.L. Landau \and H.R. Ott}
\REVIEW{J. of Low Temp. Phys.}{130}{2003}{287}.

\bibitem{Norman}
\Name{M.R. Norman and C. Pepin}
\REVIEW{cond-mat/0302347}{}{2003}{}.

\bibitem{Kastner}
\Name{M.A. Kastner  \textit{et al}.}
\REVIEW{Rev. Mod. Phys.}{70}{1998}{897}.

\bibitem{SASAGAWA}
\Name{T. Sasagawa \textit{et al}.}
\REVIEW{Phys. Rev. B}{61}{2000}{1610}.

\bibitem{Gilardi2}
\Name{R. Gilardi \textit{et al}.}
\REVIEW{Int. J. Mod. Phys. B}{17}{2003}{3411};
cond-mat/0302413.

\bibitem{GILARDI}
\Name{R. Gilardi \textit{et al}.}
\REVIEW{Phys. Rev. Lett.}{88}{2002}{217003}.

\bibitem{Cheong}
\Name{S-W. Cheong \textit{et al}.}
\REVIEW{Phys. Rev. Lett.}{67}{1991}{1791}.

\bibitem{BULUT} 
\Name{N. Bulut and D. J. Scalapino}
\REVIEW{Phys. Rev. B}{53}{1996}{5149}.

\bibitem{Norman2}
\Name{M.R. Norman}
\REVIEW{Phys. Rev. B}{61}{2000}{14751}.

\bibitem{TRANQUADA}
\Name{J. M. Tranquada \textit{et al}.}
\REVIEW{Nature}{375}{1995}{561}.

\bibitem{ZHANG}
\Name{S.-C. Zhang \textit{et al}.}
\REVIEW{Science}{275}{1997}{1089}.

\bibitem{Hu}
\Name{J.P. Hu and S.C. Zhang}
\REVIEW{cond-mat/0108273}{}{2001}{}.

\bibitem{LAKE01}
\Name{B. Lake \textit{et al}.}
\REVIEW{Science}{291}{2001}{1759}.

\bibitem{ODA}
The TSFZ-method was used for the crystal growth, see e.g. \Name{T. Nakano  \textit{et al}.}
\REVIEW{J. Phys. Soc. Jpn.}{67}{1998}{2622}.

\bibitem{YAMADA}
\Name{K. Yamada \textit{et al}.}
\REVIEW{Phys. Rev. B}{57}{1998}{6165}.

\bibitem{LEE2}
\Name{C.H. Lee \textit{et al}.}
\REVIEW{ J. Phys. Soc. Jpn.}{69}{2000}{1170}.

\bibitem{LAKE99}
\Name{B. Lake \textit{et al}.}
\REVIEW{ Nature}{400}{1999}{43}.

\bibitem{MASON}
\Name{T.E. Mason \textit{et al}.}
\REVIEW{ Phys. Rev. Lett.}{77}{1996}{1604}.

\bibitem{MORR}
\Name{D.K. Morr and D. Pines}
\REVIEW{ Phys. Rev. B}{61}{2000}{R6483}.

\bibitem{ANDO}
\Name{Y. Ando \textit{et al}.}
\REVIEW{ Phys. Rev. B}{60}{1999}{12475}.


\bibitem{GOMORY}
\Name{F. G\"om\"ory}
\REVIEW{Supercond. Sci. Technol.}{10}{1997}{523}.

\bibitem{Abrikosov}
\Name{A.A. Abrikosov}
\REVIEW{Zh. Eksp. Teor. Fiz.}{32}{1957}{1442}.

\bibitem{HAO}
\Name{Z. Hao \textit{et al}.}
\REVIEW{Phys. Rev. B}{43}{1991}{2844}.

\bibitem{OTT}
\Name{I.L. Landau and H.R. Ott}
\REVIEW{Phys. Rev. B}{66}{2002}{144506}.

\bibitem{DAnna}
\Name{G. D'Anna  \textit{et al}.}
\REVIEW{Phys. Rev. Lett.}{81}{1998}{2530}.

\bibitem{Schilling}
\Name{A. Schilling \textit{et al}.}
\REVIEW{Phys. Rev. Lett.}{78}{1997}{4833}.

\bibitem{LAKE02}
\Name{B. Lake \textit{et al}.}
\REVIEW{Nature}{415}{2002}{299}.

\end{thebibliography}
\end{document}